\documentclass[preprint,aps,pre,showpacs,notitlepage,superscriptaddress,longbibliography,floatfix]{revtex4-2}

\usepackage{hyperref}
\usepackage{graphics}
\usepackage{latexsym,exscale,stmaryrd,amssymb,amsmath}

\usepackage{tikz}
\usetikzlibrary{arrows}
\usetikzlibrary{positioning}
\usetikzlibrary{fit}
\usepackage{subfigure}
\usepackage{tabularx}
\usepackage{booktabs}
\usepackage{multirow,bigstrut}
\begin{document}
\pagenumbering{arabic}

%\title{Modulations of Patchy Particle Model}
\title{Enhanced condensate fluidity in modified patchy particle models}
\date{\today} 

\author{Alena Taskina}
\thanks{These authors contributed equally.}
\affiliation{University of G\"ottingen, Institute for the Dynamics of Complex Systems, Friedrich-Hund-Platz 1, 37077 Göttingen, Germany}
\affiliation{Max Planck School Matter to Life}

\author{Devika Magan}
\thanks{These authors contributed equally.}
\affiliation{University of G\"ottingen, Institute for the Dynamics of Complex Systems, Friedrich-Hund-Platz 1, 37077 Göttingen, Germany}
\affiliation{Indian Institute of Science Education and Research Mohali, India}
\author{Simon Dannenberg}
\thanks{These authors contributed equally.}
\affiliation{University of G\"ottingen, Institute for the Dynamics of Complex Systems, Friedrich-Hund-Platz 1, 37077 Göttingen, Germany}

\author{Stefan Klumpp}
\thanks{E-mail: simon.dannenberg@uni-goettingen.de, stefan.klumpp@phys.uni-goettingen.de}
\affiliation{University of G\"ottingen, Institute for the Dynamics of Complex Systems, Friedrich-Hund-Platz 1, 37077 Göttingen, Germany}
\affiliation{Max Planck School Matter to Life}

\begin{abstract}
Biomolecular condensates are formed via liquid-liquid phase separation of proteins, often together with nucleic acids, typically driven by interactions between low-affinity binding sites. The computational study of such condensates that accounts both for the droplet-scale fluid behavior and the internal structure of the condensate requires coarse-grained  models. Recently, patchy particle models, representing proteins as sphere with a repulsive core and directional attractive patches, have emerged as a powerful tool. However, these simulations are typically limited by slow dynamics and struggle to capture the full range of material properties of fluid-like condensates.

Here we study modified patchy particle models to simulate the formation and dynamics of biomolecular condensates. By incorporating flexible patches and weak isotropic attractions between cores, our models preserve key equilibrium characteristics, including the phase behavior and the local structure of the condensate, while significantly accelerating the system dynamics. These modifications enable the simulation of larger, more complex systems previously inaccessible due to prohibitive relaxation times and provide a versatile tool for studying condensate dynamics.
\end{abstract}

\maketitle
\newpage

\section{Introduction}
Liquid-liquid phase separation (LLPS) has emerged as a physical mechanism for cellular organization, driving the formation of membrane-less organelles and biomolecular condensates \cite{brangwynne2009germline}. These condensates play crucial roles in various cellular processes, from stress response to gene regulation, and their dysregulation has been implicated in numerous diseases \cite{banani_review,soding2020mechanisms}. Understanding the physical principles governing LLPS is therefore essential for both basic cell biology and therapeutic development. Most computational studies of LLPS have employed simple Lennard-Jones particles, which capture the basic physics of phase separation \cite{Liu_LJ_LLPS} but cannot account for the specific structural features, in particular the directional interactions characteristic of biological molecules. This limitation is particularly significant given that many biomolecular condensates are formed through specific, anisotropic interactions between proteins and nucleic acids \cite{MITTAG20222201}. These interactions are typically also not captured by thermodynamic continuum descriptions of the condensates \cite{hyman2014liquid}.

Specific molecular interactions can be captured by simulations with atomistic or amino acid-scale resolution, however at great computational cost. A coarse-grained alternative are patchy particles, which consist of a (typically spherical) core with discrete attractive interaction sites; these provide model systems for studying such directional interactions \cite{bol1982monte,Kern_Frenkel,Schluttig}. These models have been used extensively as simplified models for water-like fluids and to describe colloidal systems with directional interactions including their self-assembly and crystallization \cite{bol1982monte,bianchi2006phase,bianchi2011patchy,zhang2004self}. Patchy particles can form extended networks and undergo liquid-liquid phase separation, making them suited to describe biomolecular condensates, for which they have indeed been used in several studies \cite{Espinosa_2020,Liu_2007, nguemaha2018liquid, Quinn}. The classical patchy particle model features rigid particles with fixed interaction sites \cite{Kern_Frenkel}. While this model has provided valuable insights into
equilibrium properties of directionally-bonded systems, it faces limitations in capturing dynamic processes due to extremely slow relaxation times \cite{MarinAguilar}. 
This sluggish dynamics poses practical challenges for computational studies and limits their application to biomolecular condensate that typically require fluid-like behavior.

Here, we systematically investigate how modifications to the classical patchy particle model can overcome these challenges and result in more dynamical condensates without modifying their structural properties. 
We explore three variants of patchy particle models: (1) flexible patches that can move relative to the particle core, (2) addition of isotropic attractions between particle cores, 
and (3) a combination of both modifications. %Our goal is to identify approaches that maintain the essential structural features of patchy particle systems while enhancing their dynamic properties.
Some aspects of such modifications of patchy particle models have been studied before \cite{Palaia, smallenburg2013liquids, Espinosa2019}; these approaches will be discussed at the end of this study. 
We find that both flexible patches and weak isotropic attractions preserve the key equilibrium characteristics - including the phase behavior and the local 
structure - while accelerating the dynamics. The mechanisms for the speed-up however differ: flexible patches increase the local mobility by enlarging the accessible volume for 
bonded particles, while isotropic attractions facilitate bond breaking and reformation. Combining both modifications yields additive benefits, 
enabling up to an order of magnitude faster dynamics. % while maintaining the desired self-assembly properties.

Our results demonstrate that these modifications to the classical patchy particle model can significantly accelerate system dynamics while preserving 
essential structural characteristics. This acceleration enables the simulation of larger and more complex systems that were previously 
computationally prohibitive due to very slow  relaxation. For example, our modified models facilitate the study of polydisperse mixtures, 
relevant for biological condensates.
\section{Classical Patchy Particle Model}

%\subsection{Model Description}

We begin our investigation with the reference case of rigid patchy particles that has been used extensively in the literature \cite{Espinosa_2020,Espinosa2019}: The patchy particles are rigid bodies consisting of a spherical core with three attached patches (Fig.~\ref{fig:fig1_1}A). The three-patch design represents the minimal number of patches needed to enable network formation. Each patch is modeled as a smaller sphere fixed to the core's surface, with the patches positioned to form equal 120$^\circ$ angles with respect to each other. The cores interact repulsively through excluded volume interactions, while patches on different particles can attract each other via a Mie potential (see Methods). The entire particle moves as a rigid body, maintaining fixed relative positions between all components during the simulation.

%\subsection{Equilibrium Properties}
\subsection{Phase Diagram}

\begin{figure}[tb]
    \centering
    \includegraphics[width=1\linewidth]{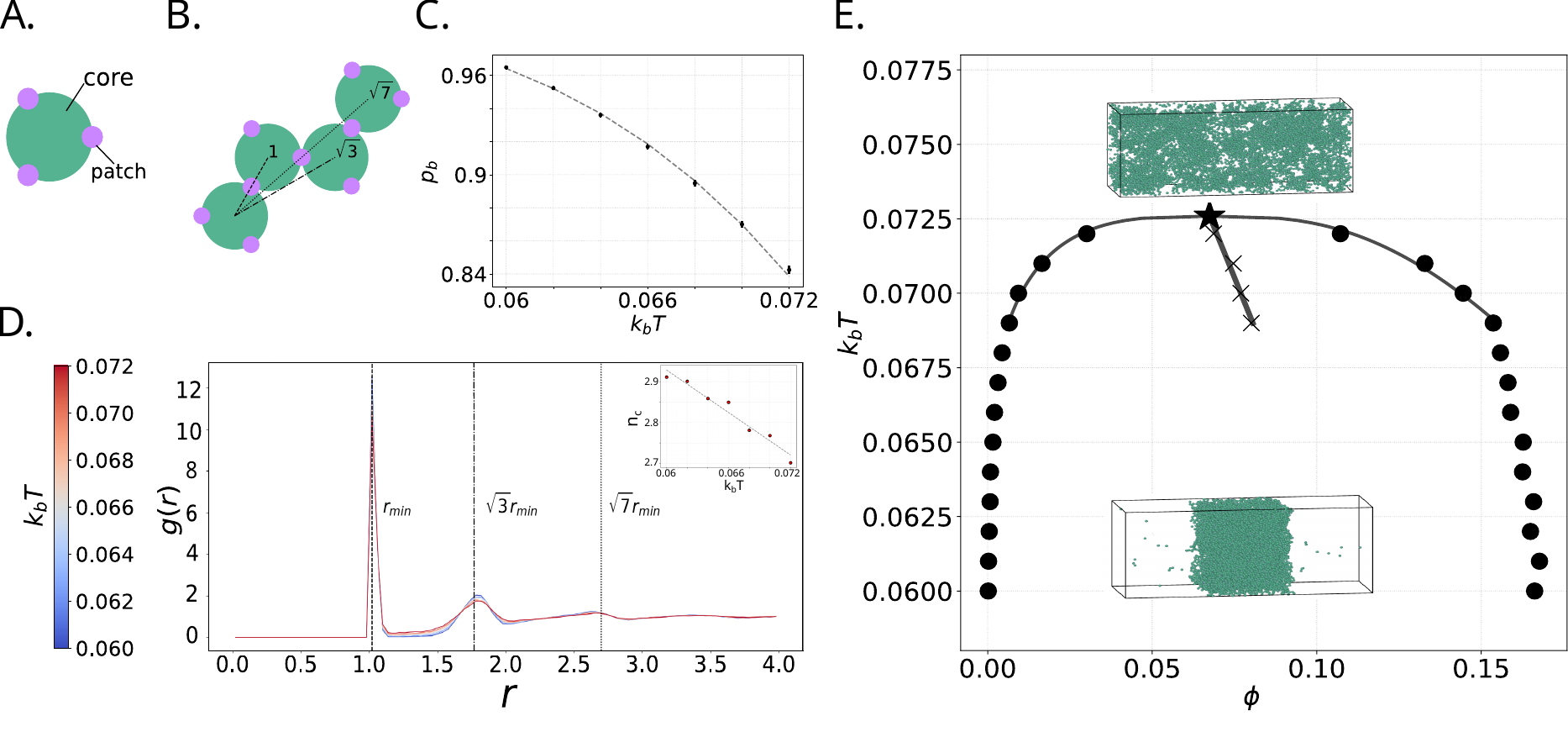}
    \caption{Phase diagram and structural properties of rigid patchy particles: A. Patchy particle B. Chain of patchy particles interacting through their patches. Distances between centers of particles are specified. C. Probability of bond formation depending on system temperature. Dashed line shows exponential fit $1 - C\exp{\frac{A}{k_bT}}$. Error bars display standard deviations. D. Radial distribution function at different temperatures. Distances between centers of particles in the chain are specified. In onset: coordination number depending on temperature. Dashed line shows linear fit. E. Phase diagram. The dots display packing fractions in dilute and condensed phases, crosses show the mean between these packing fractions, the star corresponds to critical value. The lines show the fit to the law of rectilinear diameter.}
    \label{fig:fig1_1}
\end{figure}

Our molecular dynamics simulations of 10,000 patchy particles in the NVT ensemble reveal a system that exhibits liquid-vapor phase separation. The phase behavior of our patchy particle system was mapped using the direct coexistence method (see Methods), which led to the phase diagram shown in Fig.~\ref{fig:fig1_1}E. The system exhibits different behavior across the studied temperature range from $T^* = 0.06$ to $T^* = 0.072$ (here and later $T^*$ is in dimensionless units of $k_BT/\epsilon_{attr}$ measured relative to the strength $\epsilon_{attr}$ of the attractive interaction between patches). At the lowest temperature, particles predominantly exist in a condensed phase, with negligible presence in the dilute phase. As temperature increases, the distinction between dilute and condensed phases gradually diminishes, becoming barely noticeable at $T^* = 0.072$, near the critical temperature. The critical temperature was estimated to be $T^*_c = 0.0728$ based on the law of critical exponents (see Methods).

Thus, the temperature range from $T^* = 0.06$ to $T^* = 0.072$ spans different regimes of phase behavior, from strong phase separation to near-critical conditions. In the following sections, we will explore how this temperature range affects various structural and dynamic properties of the system.

\subsection{Structural characteristics}
To better understand the microscopic basis of the condensed phase, we first analyzed two key structural properties: particle connectivity and spatial organization.

The particle connectivity was quantified using the fraction of bonded patches $p_b$. As temperature increases, bonds between particles are disrupted, a process that can be described using mean field theory (see methods). The validity of this theoretical framework is confirmed by the linear relationship observed between $\ln(1-p_b)$ and  $\frac{1}{k_BT}$ (Fig.~\ref{fig:fig1_1}C). Despite the temperature-induced bond disruption, the effect remains moderate, with $p_b$ staying above 0.85 across the studied temperature range. This high bonding fraction indicates that the condensed phase maintains a persistent network structure, even at temperatures approaching the critical point.

To gain insights into how the particles organize spatially, we characterized the system using the Radial Distribution Function (RDF), $g(r)$. The RDF exhibits three distinct peaks with progressively decreasing amplitudes at increasing distances, eventually converging to unity (Fig.~\ref{fig:fig1_1}D). This convergence indicates the liquid-like disordered state of the system. Unlike simple liquids, which show peaks at integer multiples of the particle diameter $\sigma$, our system exhibits more closely spaced peaks. This pattern suggests that particles preferentially form directional bonds via their patches, minimizing non-specific interactions with neighboring particles. Indeed the location of the peaks at $r_\text{min}$, $\sqrt{3}r_\text{min}$ and $\sqrt{7}r_\text{min}$ correspond to the next, second and third neighbor along a chain in a network (Fig.~\ref{fig:fig1_1}B).

The average coordination number ranges from 2.91 to 2.7 (Fig.~\ref{fig:fig1_1}B inset). This value remains consistently below the available number of patches per particle (three), further supporting the prevalence of directional bonding in the system.

\subsection{Dynamic Properties}
Next, we analyzed the dynamic properties of the system, such as bonds dynamics and particle mobility.
\begin{figure}[tb]
    \centering
    \includegraphics[width=1\linewidth]{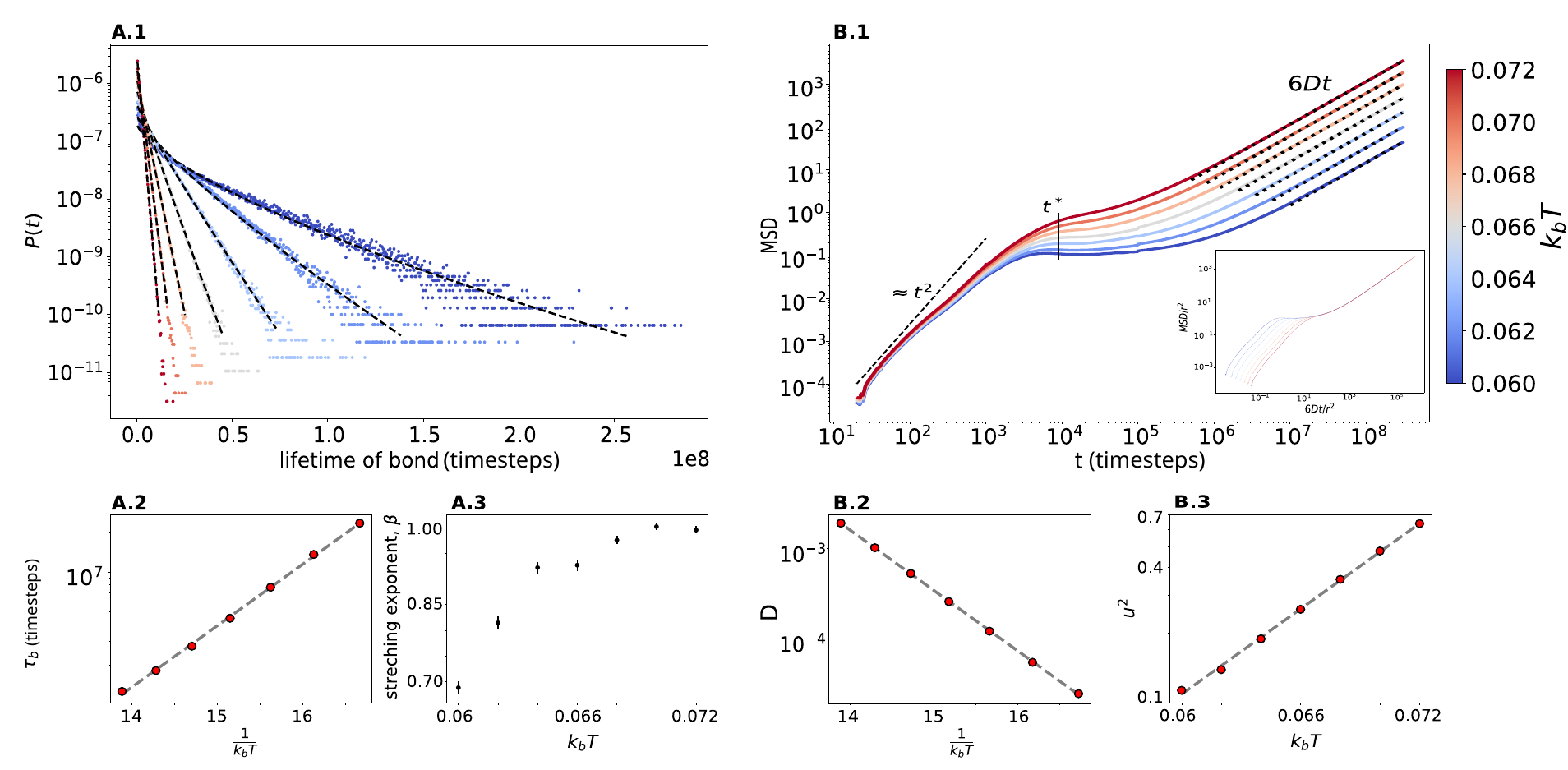}
    \caption{Dynamical properties of rigid patchy particles: A.1 Log plot of bonds lifetimes distributions at different temperatures. Dashed lines show exponential fits. A.2 Log plot of mean bond lifetime depending on reverse temperature. Dashed line shows exponential fit of Arrhenius law. A.3 Stretching exponent $\beta$ depending on temperature. Error bars display standard errors. B.1 MSD curves at different temperatures. dashed line shows the guide line for ballistic regime, dotted lines show the fits to diffusive regime. $t^*$ shows the onset of plateaus, where its' heights were measured. In onset: MSD curves form a master curve after rescaling. B.2 Log plot of diffusion coefficient depending on temperature B.3 Plateau height depending on temperature.}
    \label{fig:fig1_2}
\end{figure}
%\subsubsection{Bonds Dynamics}
To analyze bonds dynamics, we measured the lifetimes of bonds between attractive patches and determined the corresponding lifetime distribution.
The lifetime distributions follow the Kohlrausch-Williams-Watts (KWW) stretched exponential form:
\begin{equation}
\label{eq:KWW}
P(t) = A \exp\left[-\left(\frac{t}{\tau'_b}\right)^\beta\right]
\end{equation}
where $\beta$ ($0<\beta<1$) is the stretching exponent and $\tau'_b$ is a characteristic time (Fig.~\ref{fig:fig1_2}A.1). The emergence of stretched exponential relaxation suggests that our patchy particle system exhibits dynamical features characteristic of glass-forming materials. The stretching exponent $\beta$ shows a monotonic increase with temperature (Fig.~\ref{fig:fig1_2}A.3). This behavior indicates a transition from heterogeneous dynamics at low temperatures to more homogeneous, purely exponential decay at elevated temperatures.

The mean bond lifetime $\tau_b$ exhibits remarkably slow dynamics, exceeding $10^6$ simulation time steps. As shown in Fig.~\ref{fig:fig1_2}A.2, it displays an Arrhenius temperature dependence, which is a characteristic behavior of strong glass-forming materials.

%\subsubsection{Particle Mobility}

Having characterized the bond dynamics, we next examined how these microscopic bond rearrangements influence the overall particle motion within the network.
The system's dynamics were characterized by the Mean Squared Displacement (MSD) of the particle cores. The MSD curves (Fig.~\ref{fig:fig1_2} B.1) reveal three distinct regimes, again reminiscent of glass-forming systems. These include initial ballistic motion at short times, an intermediate plateau regime where particles are temporarily trapped by their neighbors, and long-time diffusive behavior.

In the intermediate plateau regime, the MSD reaches a characteristic value, the cage size %Debye-Waller factor 
$u^2$. This plateau height increases exponentially with temperature while remaining below $1\sigma^2$ (Fig.~\ref{fig:fig1_2} B.3). It indicates that despite thermal activation, particles remain locally confined by their neighbors. With decreasing temperature the plateau gets extended to longer time scales.

The long-time diffusive regime was quantified by fitting to $MSD = 6Dt$. The temperature dependence of the diffusion coefficient, as shown in Fig.~\ref{fig:fig1_2} B.2, follows the Arrhenius law, reflecting the temperature dependence of the bond lifetimes. Additionally, we found that $D$ scales linearly with the number of unbound particles $N(1-p_b)^3$. This proportionality indicates that the diffusive motion is primarily driven by freely moving, unbound particles. 

When the MSD curves are rescaled by $6D/u^2$ along the time axis and $1/u^2$ along the displacement axis, they collapse onto a master curve for times beyond the ballistic regime (inset in Fig.~\ref{fig:fig1_2} B.1). Therefore, the system demonstrates time-temperature superposition in both the plateau and diffusive regimes.

\section{Flexible Patchy Particle Model}
\begin{figure}[tb]
    \centering
    \includegraphics[width=1\linewidth]{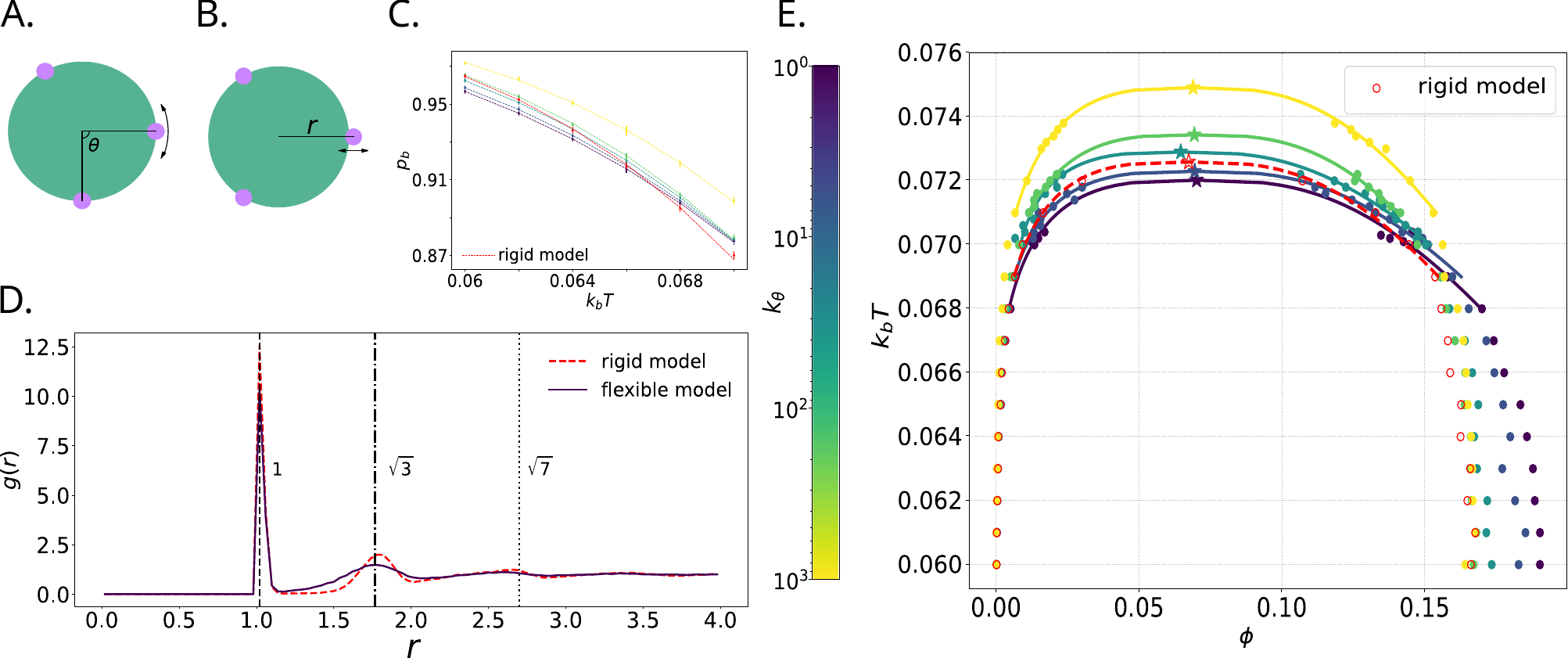}
    \caption{Phase diagram and structural properties of a system of flexible patchy particles: A. Flexible patchy particle, the arrow shows the angle change. B. Flexible patchy particle, the arrow shows the change of bond length. C. Probability of bond formation depending on system temperature for different angle stiffnesses. Dashed line shows exponential fit $1 - C\exp{\frac{A}{k_bT}}$. Error bars display standard deviations. D. Radial distribution function. Distances between centers of particles in the chain are specified. E. Phase diagrams for different angle stiffnesses. The dots display packing fractions in dilute and condensed phases, the stars correspond to critical temperatures. The lines show the fits to the law of rectilinear diameter.}
    \label{fig:fig2_1}
\end{figure}

The classical patchy particle system studied so far exhibits characteristics typical of glass-forming materials, with remarkably slow dynamics at both microscopic and macroscopic scales. This slow behavior leads to extremely long equilibration times in MD simulations, making the system computationally challenging to study. Likewise, it results in condensates that show slow dynamics and are less fluid than desired if one wants to apply patchy particle models to protein condensates.   In the following sections, we explore various modifications to the model aimed at enhancing the system's fluidity and accelerating its dynamics: we make the patchy particles flexible by allowing patches to move relative to the core, we introduce long-range isotropic interactions between particles allowing weaker interaction between patches, and finally we combining both approaches.

%\subsection{Model Description}
The first modification we explore makes the patchy particles more flexible by allowing patches to move relative to the core. This flexibility is governed by two potentials: a harmonic stretching potential controlling the core-patch distance, $V(r) = k_{b}(r - r_{0})^2$, and an angular potential regulating the angles between patches, $V(\theta) = k_{\theta}(\theta - \theta_{0})^2$ (Fig.~\ref{fig:fig2_1} A and B). The stretching potential keeps the patches near the core's surface with $k_b = 1000$. The angular potential, with variable stiffness $k_{\theta}$, allows the patches to move on the core's surface, while it also prevents patches from collapsing together and preserves the particle's three-fold symmetry with equilibrium angles of 120$^\circ$.
In the following, we will analyze how particle flexibility, controlled by the angle stiffness $k_{\theta}$, affects both equilibrium and dynamic properties of the system.

%\subsection{Equilibrium Properties}
\subsection{Phase diagram and structural properties}
The introduction of flexibility leads to significant changes in both phase behavior and structural organization of the system.
%\subsubsection{Phase Behavior}
Particles flexibility does not disrupt the ability of the system to phase separate, but affects the phase behavior of the system in two ways (Fig.~\ref{fig:fig2_1} E). First, the critical temperature decreases with decreasing angle stiffness. This indicates that more flexible particles form less stable condensates. Second, the packing fraction increases as particles become more flexible. This occurs because flexible particles can deform and pack more efficiently. Both effects are rather moderate, however, in particular compared to the modifications discussed below.

%\subsubsection{Structural Properties}
The structural organization of the system shows features that are preserved when introducing flexibility and features that are altered. The radial distribution function maintains its qualitative profile (Fig.~\ref{fig:fig2_1} D), suggesting that patch-mediated bonds still govern the overall structure. However, the bonding probability decreases with increasing flexibility, particularly at lower temperatures (Fig.~\ref{fig:fig2_1} C). This indicates that patch mobility partially inhibits bond formation.

In summary, the analysis of equilibrium properties reveals that flexibility mildly affects the density and stability of the condensed phase, while preserving its basic structural organization.

\subsection{Dynamic properties}

\begin{figure}[t]
    \centering
    \includegraphics[width=1\linewidth]{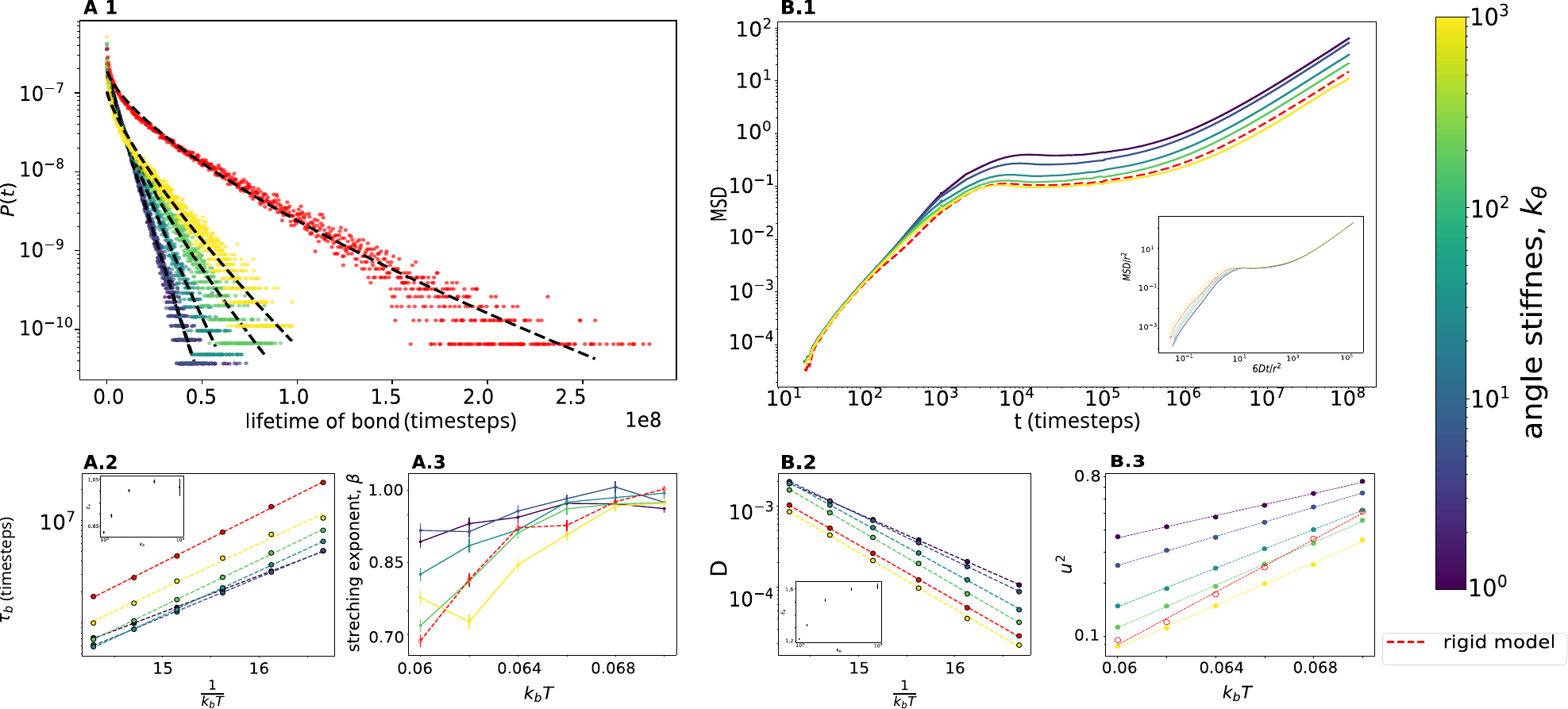}
    \caption{Dynamic properties of a system of flexible patchy particles: A.1 Log plot of bonds lifetimes distributions at $k_bT = 0.6$ for different angle stuffiness. Dashed lines show exponential fits. A.2 Log plot of mean bond lifetime depending on reverse temperature. Dashed line shows exponential fit of Arrhenius law. In onset: Activation energy depending on angle stiffness. Error bars represent standard errors. A.3 Stretching exponent $\beta$ depending on temperature for different angle stiffness. Error bars display standard errors. B.1 MSD curves at $k_bT = 0.6$ for different angle stiffness. In onset: MSD curves form a master curve after rescaling. B.2 Log plot of diffusion coefficient depending on temperature. In onset: Diffusion activation energy versus angle stiffness. Error bars represent standard errors.  B.3 Plateau height depending on temperature.}
    \label{fig:fig2_2}
\end{figure}

We now examine how flexibility influences both bond dynamics and particle mobility.
%\subsubsection{Bond Dynamics}
Bond lifetime distributions maintain their KWW form (Fig.~\ref{fig:fig2_2} A.1), revealing persistent glass-like dynamics. Systems with lower angle stiffness show larger $\beta$ values, approaching 1 even at lower temperatures (Fig.~\ref{fig:fig2_2} A.3). This indicates that increased patch mobility leads to more uniform bond dynamics  even at low temperature, making the system more liquid-like.
Bond lifetimes remain long, exceeding $5 \cdot 10^5$ timesteps across all conditions. They follow Arrhenius temperature dependence, but with reduced activation energy when patches are more mobile (Fig.~\ref{fig:fig2_2} A.2). This shows that lateral patch movement accelerates bond dynamics, especially at lower temperatures.
%\subsubsection{Particle Dynamics}

The MSD curves qualitatively show the same behavior as for rigid patchy particles and in particular still display the characteristic plateaus, indicating temporary particle localization (Fig.~\ref{fig:fig2_2} B.1). With increased patch mobility, the particles can explore larger local volumes, as shown by higher plateau values (Fig.~\ref{fig:fig2_2} B.3). However, the duration of this localized motion remains constant, i.e.\ independent of the degree of patch flexibility.
The diffusion coefficient increases slightly with patch mobility, but stays below $8\cdot10^{-2} \frac{\sigma^2}{\tau}$ for all simulated cases. It follows Arrhenius behavior, but systems with more mobile patches exhibit lower activation energies (Fig.~\ref{fig:fig2_2} B.2). This further confirms that intramolecular dynamics mimicked by patch flexibility accelerates the system's dynamics.

Overall, the system maintains its fundamental characteristics of a strong glass-forming material, although with accelerated dynamics.
\section{Patchy Particle Model with Isotropic Potential}
%\subsection{Model Description}

As an alternative to the flexible patches, we studied another variant of the patchy particle model focused on increasing the fluidity of the system.
In this model, we introduced isotropic attractive interactions between particle cores, in addition to the existing repulsive interactions. By adding an attractive interaction with strength $\epsilon_{\text{core}}$ between the cores, we can in turn weaken the interaction strength of the patches, $\epsilon_{\text{patch }}$, without dissociating the particles. This modification is expected to enhance the system’s dynamical properties by facilitating more fluid-like inter-particle movements and interactions.
The isotropic attraction allows the particles to break bonds with energy $\epsilon_{\text{patch}}$ and locally rearrange  within the condensate. %without escaping the condensate. 
While $\epsilon_{\text{patch}}$ facilitates the formation of specific bonds that dictate the structure and dynamics within
the condensate, the introduction of $\epsilon_{\text{core}}$ serves as an additional cohesive force that
maintains a general structure of the condensate.

\begin{figure}[tb]
    \centering
    \includegraphics[width=1\linewidth]{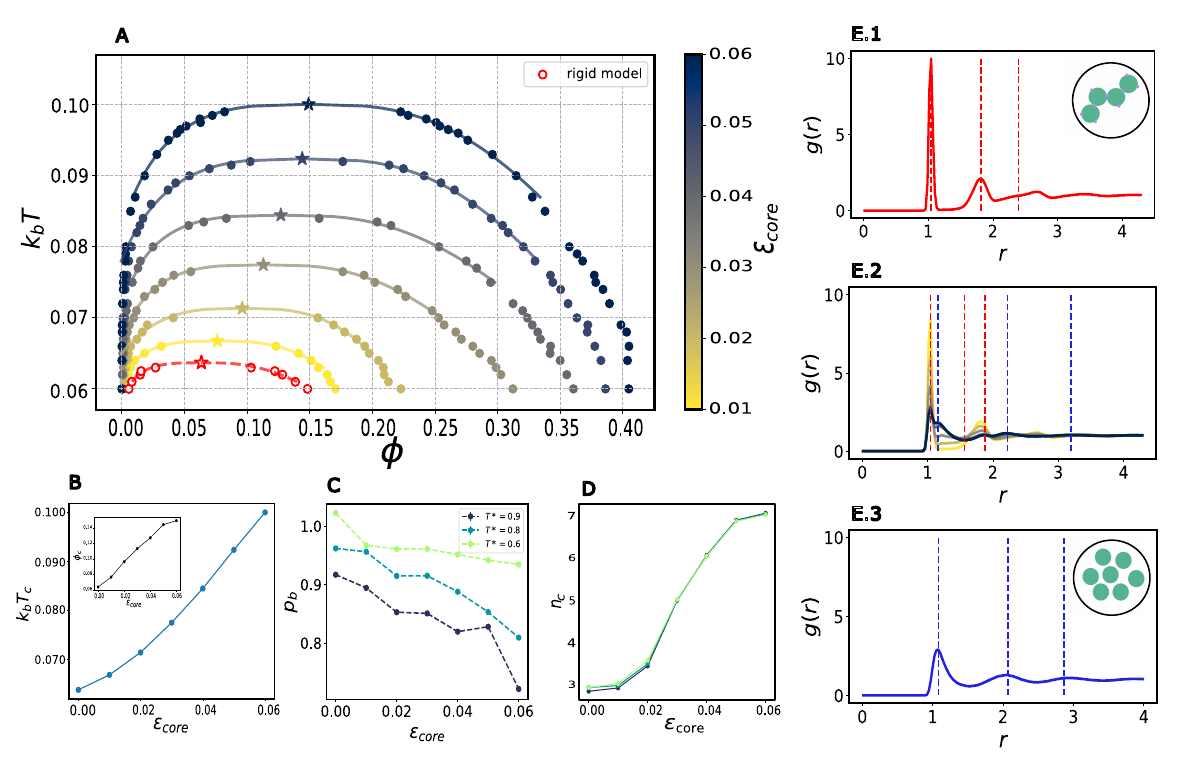}
    \caption{Phase diagram and structural properties of a system of patchy particles with an isotropic attraction: A. Phase diagrams for different core potential $\epsilon_{\text{core}}$. B. The critical temperature $k_{\text{b}}T_{\text{c}}$ as a function of $\epsilon_{\text{core}}$, indicating an increase in $T_{\text{c}}$ with increasing $\epsilon_{\text{core}}$. Inset: Critical packing fraction, $\phi_{\text{c}}$ versus $\epsilon_{\text{core}}$. C. Probability of bond formation, $p_{\text{b}}$ as a function of $\epsilon_{\text{core}}$ at different reduced temperatures, $T^*= T/T_c$. Error bars display standard deviations. D. Coordination number $n_{\text{c}}$ as a function of $\epsilon_{\text{core}}$ at different $T^*$ values. E. Radial distribution functions,  $g(r)$ for different systems: E.1.the rigid model, E.2.varying $\epsilon_{\text{core}}$, and E.3. Lennard-Jones system. Vertical dashed lines indicate peak positions corresponding to characteristic interparticle distances. Insets depict schematic representations of corresponding particle arrangements.}
     \label{fig:fig5}
\end{figure}

\subsection{Phase diagram and structural properties}
%\subsection{Liquid-vapor coexistence curve}
Fig.\ \ref{fig:fig5}A shows the phase diagram for the reference case of particles with interaction purely between patches (rigid model, red) and for different values of the strength of the attractive core potential. 
As the core potential strength increases, there is a visible shift in the coexistence curves to higher density and to higher temperature (note that here the increased core interaction is not compensated by a reduced patch interaction). The shift to higher density indicates a change in the packing of the particles in the condensate. The isotropic potential facilitates a closer packing of particles by increasing the overall attractive forces in the system, leading to a more compact condensate even at higher temperatures. Our results indicate that the isotropic potential strengthens the interactions that favor the formation of denser and more stable condensates.

%\subsection{Structure}
%\subsubsection{Radial Distribution Function}

At low values of $\epsilon_{\text{core}}$ with $\epsilon_{\text{core}} < 0.03$, the RDF calculated for the isotropic model resembles that of the rigid model system (Fig.~\ref{fig:fig5}E.1) which exhibits a sharp peak in the beginning and subsequent diminishing peaks at the well-defined short ranges observed above.
As we increase the isotropic potential strength, $\epsilon_{\text{core}} \geq 0.03$, the peaks of the RDF become less pronounced and we observe additional peaks corresponding to the RDF for a Lennard-Jones fluid (Fig.~\ref{fig:fig5}E.3, E.2). This indicates the formation of a second type of structure within the system. While the primary peaks corresponds to the original patch interactions, the secondary peaks are due to the additional layer of attraction imposed by the isotropic potential. This novel structure indicates the transition to a more complex phase with both short-range order due to particle chains mediated by patch interactions ($\epsilon_{\text{patch}}$) and medium range correlations facilitated by the isotropic attractive interactions ($\epsilon_{\text{core}}$).

%\subsubsection{Network connectivity}
Correspondingly, the network connectivity quantified by the fraction of bonded patches, $p_b$ decreases  with increasing $\epsilon_{\text{core}}$ (Fig.~\ref{fig:fig5}C), which can be attributed to the enhanced ability of particles to rearrange due to the increased isotropic interaction, coupled with the diminishing relative importance of anisotropic interactions in determining the system’s structure.

%\subsubsection{Coordination Number}
As the fraction of bonds in the system decreases, the coordination number increases with the strength of the isotropic potential (Fig.~\ref{fig:fig5}D). %A higher core leads to increased attractive forces between particles, drawing them closer together. While it doesn’t result in directional bonding, it does mean that each particle, on an average, is surrounded by more neighbors reflected by the increase in the CN.
For low $\epsilon_{\text{core}} < 0.03$, the coordination number remains in the range of 2.9 to 3.5, not deviating much from the the value expected for interactions mediated predominantly by the patches, consistent with the observation from the RDF. But as we increase the interaction strength further, the coordination number increases rather abruptly, up to 7.2 for $\epsilon_{\text{core}} = 0.06$. The sudden transition suggests a threshold beyond which the isotropic potential becomes dominant and significantly alters the configuration of the particles, leading to a more densely packed system, again consistent with the observed changes in the RDF.  
%This could also be indicative of a crossover from one type of structural organization to another.

\subsection{Dynamics}
%\subsubsection{Bonds persistence}
%\subsubsection{Mobility and Diffusion}

\begin{figure}[h]
    \centering
    \includegraphics[width=1\linewidth]{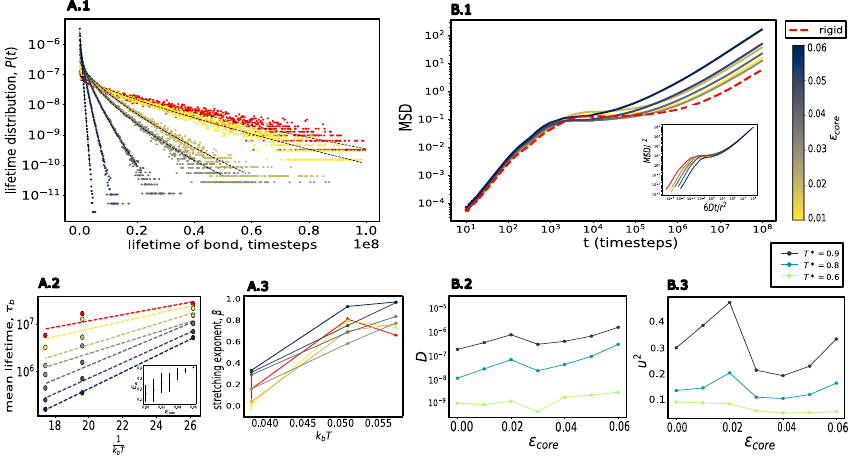}
    \caption{Dynamic properties of a system of patchy particles with an isotropic attraction: A.1. Log plot of bonds lifetime distributions at $T^* = 0.8$ for different values of $\epsilon_{\text{core}}$. Dashed lines show exponential fits. A.2. Log plot of mean bond lifetime depending on reverse temperature. Dashed lines show exponential fit of Arrhenius law. Inset: Activation energy depending on $\epsilon_{\text{core}}$. Error bars represent standard errors. A.3. Stretching exponent, $\beta$ depending on temperature for different $\epsilon_{\text{core}}$. Error bars display standard errors. B.1. MSD curves for different $\epsilon_{\text{core}}$ at $T^* = 0.8$. Inset: MSD master curves. B.2. Diffusion coefficient plotted against $\epsilon_{\text{core}}$ at various values of $T^*$. B.3. Plateau height as a function of $\epsilon_{\text{core}}$ at different values of $T^*$. }
    \label{fig:fig6}
\end{figure}

The isotropic model, compared to the rigid system, exhibits faster bond dynamics while maintaining key similarities in qualitative features. The bond lifetime distributions exhibit a stretched exponential form across different values of $\epsilon_{\text{core}}$ (Fig.~\ref{fig:fig6}A.1). When analyzing mean bond lifetimes $\tau_b$, we observe a clear Arrhenius dependence (Fig.~\ref{fig:fig6}A.2), but with a significantly reduced activation energy compared to the rigid model. This reduction leads to markedly shorter bond lifetimes, suggesting enhanced bond dynamics.
The stretching exponent, $\beta$ approaches unity more rapidly in the isotropic model, indicating a transition to uniform bond dynamics for high values of $\epsilon_{\text{core}}$ (Fig.~\ref{fig:fig6}A.3).

 For the isotropic model, the MSD curves also reflect increased particle mobility in the system as indicated by shorter plateaus and a more pronounced diffusive regime (Fig.~\ref{fig:fig6}B.1). Across different values of $\epsilon_{\text{core}}$, the curve is similar to that of the rigid model but upon increasing the potential strength, the trend is not monotonic. For $\epsilon_{\text{core}} < 0.03$, the plateau height of the MSD increases indicating a shorter plateau as well as a greater spatial confinement increasing particle mobility. However, when $\epsilon_{\text{core}} \geq 0.03$, the system shows an interesting behavior: while the plateau shortens, indicating quicker transitions between states, the overall MSD height diminishes slightly, suggesting tighter spatial confinement within the newly formed dynamic structures.

\section{Flexible Patches combined with an Isotropic potential}

\begin{figure}[tb]
    \centering  
    \includegraphics[width=1\linewidth]{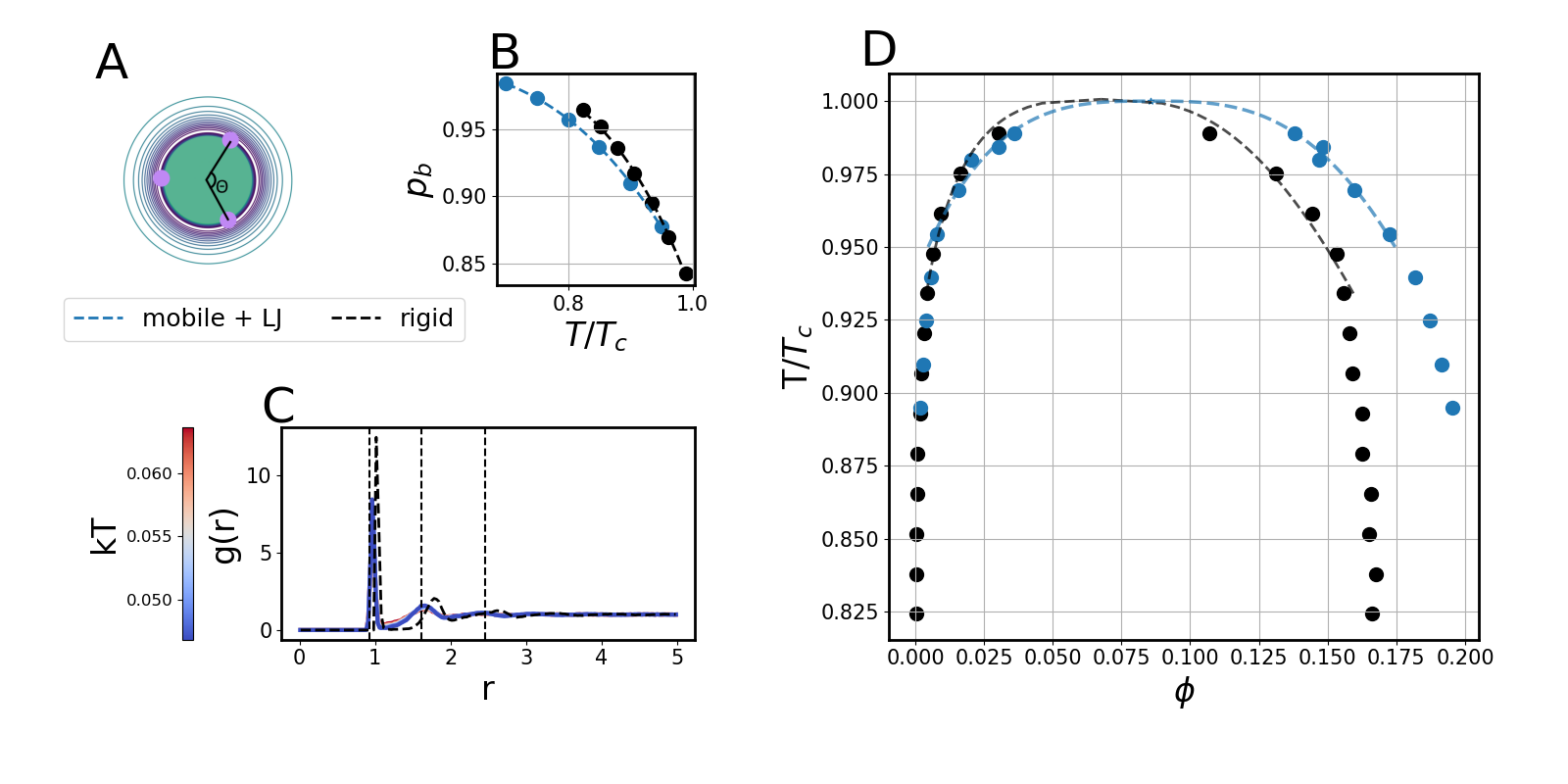} 
    \caption{Phase diagram and structural properties of patchy particle systems with flexible pacthes and an isotropic potential: \textbf{A} Schematic illustration of the combination of mobile patches with an Isotropic potential \textbf{B} fraction of formed bonds in condensate \textbf{C} Radial distribution function of particles in condensate \textbf{D} Phasediagram of Mobile patchy particles with weak isotropic core potential.} 
    \label{structure_combination_model}
\end{figure}

To maximize the mobility of particles inside the condensate while maintaining its structure, we finally combine the two modifications of the patchy particle model that we have seen to make the patchy particles system more fluid.  %a combination of the two prior models. 
For this, we add both mobile patches as well as a weak isotropic potential to our patchy particle model (schematic depiction in Fig. \ref{structure_combination_model}a), following the approaches described above. We use a spring stiffness of $k = 1$ and a isotropic potential strength of $\epsilon = 0.01$, as we have seen that these parameter choices individually cause negligible change to the radial distribution functions, phase diagrams and bond formation of the network, while increasing the mobility of the particles.

Analyzing the mentioned structural quantities for the model with both modifications, we see little effect compared to the rigid model: The fraction of bonds formed remains very high as expected for a fully bonded network (Fig. \ref{structure_combination_model}B) and only deviates slightly from the rigid model (black). Further, the radial distribution function stays qualitatively the same as in the rigid case, however the peak positions are less localised due to the stiffer potential. (Fig. \ref{structure_combination_model}c). The phase diagram shows slightly higher densities in the condensate,  similar to the changes already observed in the case of a rigid patchy particle with an added isotropic potential. Also the critical temperature changes to $T_c = 0.067$, which is slightly lower compared to $T_c^{rigid} = 0.0728$ in the rigid case (Fig.~\ref{structure_combination_model}D~).

\begin{figure}[tb]
    \centering  
    \includegraphics[width=1\linewidth]{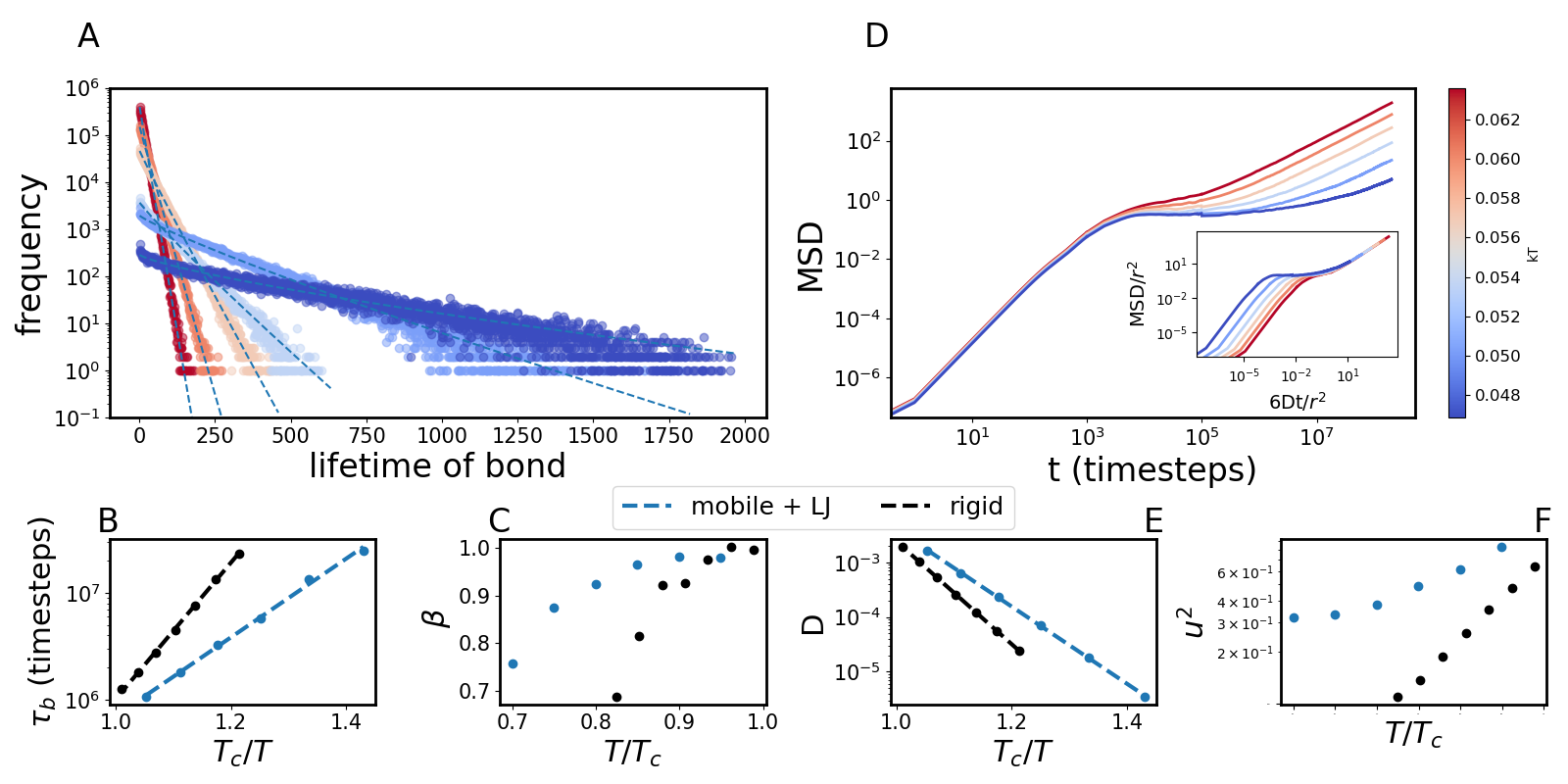}   
    \caption{Dynamics of patchy particles with with weak isotropic interactions and flexible patches: \textbf{A} Lifetime distributions of bonds  fitted to a stretched exponential for different temperatures. \textbf{B} \& \textbf{C} obtained fitted lifetime of the bonds $\tau_b$ and exponent $\beta$. \textbf{D} MSD curves for different temperatures. \textbf{E} \& \textbf{F} obtained diffusion coefficients and plateau heights.}
    \label{Bonds_combination_model}
\end{figure}
The dynamical characteristics of the combined model are also qualitatively similar to the rigid model, but the dynamics is faster. The bond lifetime distribution display the same qualitative features as the rigid model and follows the shape of a stretched exponential (Fig\ref{Bonds_combination_model}A).  For a quantitative comparison to the rigid model, we obtained the mean lifetime $\tau_b$  and the stretching exponent $\beta$. We plotted these quantities over the ratio of $T_c/T$ and $T/T_c$, respectively, and show also the values obtained for the rigid model (in black) in Fig.~\ref{Bonds_combination_model}B\&C.  Notably, our modifications do not affect the exponential Arrhenius law dependency. However, the activation energy is reduced considerably, resulting in a drastic decrease in $\tau_b$ of up to an order of magnitude. The stretching exponent also shows a qualitatively similar trend, approaching 1 for values close to the critical temperature. However, this process is started at much lower temperatures than for the rigid model. This is reflected in lifetime distributions approaching Poisson statistics for a larger range of temperatures compared to the rigid model.  
The lowered bond lifetimes result in more rapid dynamics of the system as reflected by the MSD curves shown in Fig.~\ref{Bonds_combination_model}D. Qualitatively, they share the dynamics of a super cooled liquid approaching the glass transition as in the case of the rigid model. Characterizing diffusion quantitatively  by the diffusion coefficient and plateau height as functions of the temperature (Fig \ref{Bonds_combination_model}E and F), we observe again the adherence to the Arrhenius law. The diffusion coefficients, however, are up to an order of magnitude higher than for the rigid model, showing the desired increase in mobility.
Hence, our analysis of the structural and dynamical behaviour of a patchy particle model with a weak isotropic potential and mobile patches maintains the qualitative features of the standard rigid model. However, the dynamics of the system are much faster.

\section{Discussion}

In this work we investigated how mobile patches and/or additional isotropic interactions can be introduced in a patchy particle model to enhance the dynamics of the model and to fluidize the condensates formed by liquid-liquid phase separation. We found that only in the case of strong isotropic interactions, the structural parameters are changed qualitatively. Neither mobile patches nor weak isotropic interactions between the cores result in significant alterations of the phase diagram, the radial distribution function, or the network behavior of the system. Importantly however, both speed up the dynamics: Both modifications result in shorter bond lifetimes and faster diffusion, by allowing particles to escape their cage more easily or by making the cages larger and more flexible as we will show next.

%\section{Comparision}
\begin{figure}[t]
    \centering  
    \includegraphics[width=1\linewidth]{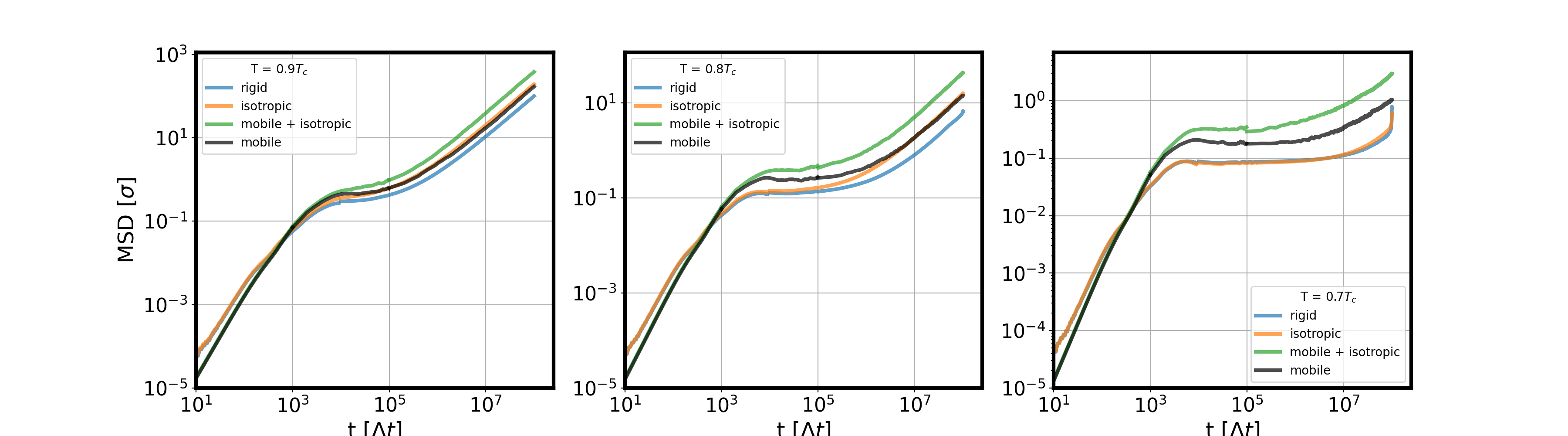}   
    \caption{Comparison of the particle mobility in the four models simulated in this study: Mean square displacement of the particles as a function of time for the four models at three different values of $T/T_c$, i.e., temperatures  relative to the respective critical temperature of the model. }
    \label{MSD_comparision}
\end{figure}

To compare the extent of the speed-up, we compared MSD curves for the different models at the same temperature relative to the respective critical temperature ($T/T_c$). The rationale for comparing the different systems at the same $T/T_c$ is that we are considering the critical temperature (here measured in units of the patch interactions strength) as a measure of the combined strength of both types of attractive interactions, so at the same $T/T_c$, an increased strength of the core potential is effectively compensated by a reduced patch attraction strength. The comparison is plotted  in Fig.~ \ref{MSD_comparision}. This plot shows that the two different modifications of the  alter the dynamics in different ways. While the mobile patches increase the height of the plateau and therefore the cage size, the isotropic potential shortens the plateau, i.e., it causes an earlier escape from said cage, while not altering its size. Combining the two modifications of the patchy particle model, the two effects appear to be additive and therefore increase the fluidity of the condensate substantially.

This can be see most drastically at  $T = 0.7 T_c$ as the plateau of the MSD is very prolonged in the case of the rigid, mobile and isotropic models. Even after $ 10^{8} \, \Delta t$ particles barely moved, making simulations at such low temperatures extremely costly with respect to computation times. Only combining both modifications of the model allows for particles to travel significant distances during that time. The same effect, though less pronounced, is also seen at higher temperatures, where the combination of the weak isotropic potential and the mobile patches increases diffusion by factors of 10 and 5 for $T = 0.8 T_c$ and $T = 0.9 T_c$, respectively. 

Several previous studies have considered similar modifications of patchy particle models.
Palaia et al. \cite{Palaia} studied the effect of a weak isotropic potential  by adding a squared cosine potential as an additional short range interaction in a two-dimensional patchy particle system. They reported increases of up to two orders of magnitude in the diffusion coefficient when balancing a decrease of the patch interaction with their isotropic potential. They did neither report how this modification affects the phase diagram of liquid-liquid phase separation nor whether affects the internal structure of the condensates as their focus was on size control of clusters. We expect that both features are influenced in their scenario, as they considered relatively strong isotropic interactions with ratios of the isotropic and patch interactions strength of up to 1/4. 

With respect to mobile patches, the studies of Espinosa et al \cite{Espinosa2019} and of Smallenburg et al. \cite{smallenburg2013liquids} introduced qualitatively similar modifications as we did. In the work of Espinosa et al., an angular potential was introduced to investigate the influence of flexible patch positioning on the phase diagram. Similar to our results, they reported a small increase of the density inside the condensate, but they reported identical critical temperatures, in contrast to the increase that we see in our results. This difference may be attributed either to the smaller system sizes used in their study or to the fact that Espinosa et al. did not simulate temperatures in the vicinity of the critical temperature. They did not address the mobility of he particles and thus the fluidity of the condensate, but we expect that their simulations show similarly increased particle mobility as ours. 

Smallenburg et al. \cite{smallenburg2013liquids} 
simulated patchy particles interacting via the Kern-Frankel potential, restricting simultaneous interactions to one per patch. For rather large patches, this also provides some mobility to the bonds similar to our case here. They reported an increase in mobility comparable to what we found, by measuring the diffusion coefficients for different patch sizes. However, they reported increased stability of the phase-separated state with increasing patch size, in contrast to our results. We suspect that this difference results from the fact that, in our model, particles in the condensate suffer a larger entropic loss, as the patches in our model have degrees of freedom when no bonds are formed. are actual moving particles. %This effect counters the increased number of configurations in which binding events are possible. 
Thus subtle details of these models likely affect their behavior.

Importantly, the two modifications of the patchy particles studied here enable the simulation of more complex systems for which equilibration typically takes very long and in which, only narrow temperature regimes would be accessible in which a condensate with high fluidity is observed. With the increased fluidity of the modified patchy particles, this temperature window widens, allowing more systematic studies. For example one can used these models to study  polydisperse mixtures in which a large patchy particle is recruited into a condensate of smaller ones. This opens up investigations of passive microcrheology in such simulations, where a larger particle is used as a tracer and it can be used for the study of biological condensates that are typically composed of multiple components, often proteins and RNA \cite{banani_review}. Another interesting example is the synaptic vesicle cluster, which was shown to be a liquid condensate. Here, the intrinsically disorded regions of the protein Synapsin 1 form the scaffold of a condensate, which contains the much larger synaptic vesicles \cite{HOFFMANN2021166961}. The latter are covered with proteins that carry additional intrinsically disordered regions that introduce complex interactions between different constituents. For such system patchy particle models provide a mesoscale description that in contrast to continuum droplet models can describe some of the internal molecular organization of the condensate without the need for atomistic detail. The speed-up of patchy particle simulations by the modifications studied here thus established a methods that is relatively easy to implement and allows the simulation of such system at reduced computational cost.

\section{Methods}

\subsection{Molecular dynamics and direct coexistence method}
To simulate the dynamics of the patchy particles we use the simulation toolkit HOOMD-blue \cite{HOOMD}. In all simulations the dynamics were calculated using the symplectic Martyna-Tobias-Klein equations of motion for the NPT and NVT ensembles, respectively \cite{Martyna_NPT_NVT}. For each setup at least 10000 patchy particles have been simulated to reduce finite size effects. The time step was chosen as $\Delta t = 1$ fs  (0.001 in HOOMD units) and the time constants of the thermostat and barostat were set to be $\tau_T = 0.2$ ps and $\tau_s = 2$ ps. Because the strength of patch potential sets the energy scale, we use dimensionless temperatures $T^* = k_b T / \epsilon_{patch}$. Similarly we set the diameter of our rigid particle $\sigma = 1 nm$ and use it as our unit of length.

To generate phase separated configurations we utilize the direct coexistence method \cite{LADD_DC}, similar to the procedure described in ref.~\cite{Espinosa2019}. For this, we first equilibrate our system in a cubic box in a NVT ensemble at a low temperature for at least $10^7$ time steps. Subsequently we switched to a NPT  ensemble at $P = 0$. Throughout the procedure, we track the number of bonds formed, the volume and potential energy to check the equilibration.

Once the system is equilibrated, we add two empty simulation boxes at opposing sides of the box,  mimicking the vapor phase. With this we generate an elongated box as depicted in the insets in FIG \ref{fig:fig1_1} E and continue the simulation in the NVT ensemble, in which the system is subsequently equilibrated for different temperatures to obtain $\rho_{gas} $ and $\rho_{liquid}$, the densities in the two coexisting phases (see below).

\subsection{Rigid patchy particles}
For the creation of rigid patchy particles \cite{Espinosa2019} we employ the rigid constraint in HOOMD-blue \cite{HOOMD}. We use a central particle as a core. Subsequently, we add three constituent particles, the so called patches. They are placed equidistantly on the equator of the core separated by a $120^\circ$ angle. We then calculate the inertia tensor of the resulting rigid body and the resulting matrix is diagonalized. The obtained eigenvectors are then used to create a reference frame of the constituent particle positions in which the inertia tensor is diagonalized (for a detailed description how to create a rigid body in HOOMD-blue, see the HOOMD documentation and tutorials, https://hoomd-blue.readthedocs.io/en/v4.9.1/index.html ). 

The patchy particles can interact with each other via steric repulsion of their cores   
and via attraction of the patches.   
The potentials, both inspired by ref. \cite{Espinosa_2020} are described below. We set the potential parameters such that only a pairs of patch can interact attractively and steric repulsion prevents three-patch interactions.

\subsubsection{Steric repulsion}
\label{core_repulsion}
We approximate the cores of patchy particles as pseudo-hard spheres \cite{jover2012pseudo} with a diameter of \(\sigma = 1\) nm and mass $m=1$ atomic unit. For this we use a Mie potential \cite{mie1903kinetischen} with exponents $n_1=50$ and $n_2=49$, which we truncate following the Weeks-Chandler-Anderson recipe with a cutoff at $r_0$ and $V'(r)=V(r)+V(r_0)$, so that the potential goes to zero at the cutoff \cite{weeks1971role}.  
This results in the following form of the potential:
\begin{equation}
    \label{eq:rep_pot}
    V_{rep}(r) = 
    \begin{cases} 
    50 \left( \frac{50}{49} \right)^{49} \epsilon_{rep} \left[ \left( \frac{\sigma}{r} \right)^{50} - \left( \frac{\sigma}{r} \right)^{49} \right] + \epsilon_{rep}, & \text{if } r < \frac{50}{49} \sigma \\
    0, & \text{if } r \ge \frac{50}{49} \sigma. 
    \end{cases} 
\end{equation}
Here, $r$ describes the distance of the cores of the rigid bodies and $\epsilon_{rep} = 1$.

\subsubsection{Patch attraction}
\label{patch_attraction}
For the patch interaction we deviate from the approach used in previous work \cite{Espinosa2019} and use a softer Mie potential instead. Specifically, we use:
\begin{equation}
    \label{eq:rep_pot}
    V_{att}(r) = 
    \begin{cases} 
    4 \epsilon_{patch} \left[ \left( \frac{2 r_{patch}}{r - \Delta} \right)^{2} - \left( \frac{2 r_{patch}}{r- \Delta} \right)^{1} \right] + \epsilon_{shift}, & \text{if } r < 2 \, r_{patch} \\
    0, & \text{if } r \ge 2 \, r_{patch}. 
    \end{cases} 
\end{equation}
In this expression, $\epsilon_{patch}= 9 $ is the attraction strength between the patches and $r_{patch} = 0.12 \sigma$ is the size of the patches.  
We choose  $\epsilon_{rep} = - V(2 r_{patch})$ and $\Delta = 4 r_{patch} $ to ensure a continuous potential at the cutoff and minimal energy at $r = 0$. The values result in a minimal potential energy of -1.

\subsection{Patchy particles with flexible patches}
\label{Methods_Phase_diagram}
For the model with flexible patches, we represent the patches as particles attached to the core with springs using a harmonic potential $V(r) = k_{b}(r - r_{0})^2$, using the implementation available in HOOMD-blue. We choose $r_{0}$ to be $\sigma /2$ and use a spring constant of $k_{b} = 1000$. This results in the patches being able to move over the surface of the core. Higher $k_{b}$ values cause numerical instabilities with the employed time step of $\Delta t = 0.001$. 

In addition, we added an angular potential between the connecting vectors of the patches and the core. Here, we choose again a harmonic potential $V(\theta) = k_{\theta}(\theta - \theta_{0})^2$ using the implementation available in HOOMD-blue. We set $\theta_{0} = 120 ^\circ$. We used different values for $k_{\theta}$ to tune the mobility of the patches.

Apart from allowing for a smooth transition between the rigid and the mobile implementation of the model, attaching patches by springs rather than making them freely mobile on the surface of the core also has other benefits: It circumvents 
unwanted interactions 
such as the mutual attraction between patches on one core particle , which results in them collapsing in one position (the same may happen if the patches interact with the same patch on a neighbouring core particle).

\subsection{Patchy particles with isotropic potential}

For patchy particles with an isotropic attractive core potential, we use a Lennard Jones potential between cores with the following form.
\begin{equation}
  F(\Vec{r}) =
    \begin{cases}
      \epsilon_{rep} \big  \{ (\frac{\sigma}{r})^{12} -   (\frac{\sigma}{r})^6 \}\, &  \,\text{$r < 2 ^{1/6}$}  \sigma \\
      \epsilon_{att} \big  \{ (\frac{\sigma}{r})^{12} -   (\frac{\sigma}{r})^6 \} \, & \, \text{ $r > 2^{1/6}$} \sigma  \, \& \,   r < r_{cut} \\
      0 & \text{else}.
    \end{cases}      
    \label{isotropic_equation}
\end{equation}
By using two different $\epsilon$ values for the attractive and the repulsive part of the potential, we assure that the repulsive part of the potential remains the same when we vary the strength of attraction. We implemented this in HOOMD-blue by using two Lennard-Jones potentials: One with $r_{cut} = 2 ^{1/6} \sigma$ and $\epsilon = \epsilon_{rep} - \epsilon_{att}$ and one with $r_{cut} = 2.5 \sigma$ and $\epsilon = \epsilon_{att}$. The resulting potential is of the form of equation \ref{isotropic_equation}.

\subsection{Analysis of simulation data}
\subsubsection{Phase Diagrams}
To obtain the densities of the condensed and dilute phase in order to construct the phase diagrams, we generate density histograms projected along the long axis (x-axis) of the simulation box for at least 100 uncorrelated simulation frames and average over them. Subsequently, we determine the position of the condensate by calculating the middle position between the two points of strongest density changes. In a window of roughly 10 $\sigma$ we average the density to obtain $\rho_{liquid}$. For $\rho_{gas}$ we do the same but calculate the middle position in the other direction crossing the periodic boundary. 
This procedure is repeated for multiple temperatures and the critical temperature is determined by fitting the critical scaling law \cite{Petsev_scaling,Broide_scaling}
\begin{equation}
   \Big ( \rho_{liquid}(T) - \rho_{gas}(T) \Big )^{3.06} = d \left(1 -\frac{T}{T_c}\right)
    %\label{isotropic_equation}
\end{equation}
to the data of $\rho_{liquid}$ and $\rho_{gas}$,  
further assuming that close to the transition the law of rectilinear diameter 
\begin{equation}
   \Big(\rho_{liquid}(T) + \rho_{gas}(T)\Big) /2 = \rho_{c} + s_2 (T_c - T)
    %\label{isotropic_equation}
\end{equation}
holds. In these expressions $d$ and $s_2$ are used as fitting parameters. 

\subsubsection{MSD, RDF, and number of bonds}
For the calculation of the number of bonds, the radial distribution function (RDF) and the mean squared displacement (MSD) we used the software freud \cite{FREUD} to analyze the trajectories obtained with HOOMD-blue. In the case of the MSD and RSD, we used the provided functions of the package directly. To quantify the number of bonds we used freud to calculate neighbour lists and counted those type pairs as bonded for which the patches are in range of their interaction potential. 
To avoid including particles leaving the condensate into the calculation, we performed NVT simulations in periodic boxes. They were done at densities of the lowest available temperature from the phase diagrams for the respective systems.

\subsubsection{Mean Field Theory of Bond Formation}
\label{Mean_Field_bonds}

The bond formation process can be described using mean field theory, where bonds form between free patches according to $F + F \rightleftharpoons B$.
For $N$ particles with $m$ patches each, $p_b = \frac{n_b}{m \cdot N / 2}$, where $n_b$ is the number of bonds.
At equilibrium, the balance between bond formation and breaking rates yields
\begin{equation}
\frac{p_b}{(1-p_b)^2} = \exp\left(-\frac{\Delta F_b}{k_bT} \right),
\end{equation}
where $\Delta F_b$ is the free energy difference per bond formation. Given the high bonding fraction observed ($p_b \approx 1$), this expression can be approximated as
\begin{equation}
1-p_b \approx \exp{\left(\frac{1}{2}\frac{\Delta F_b}{k_bT}\right)}.
\end{equation}
The validity of this theoretical framework is confirmed by the linear relationship observed between $\ln(1-p_b)$ and $\frac{1}{k_bT}$ (Fig.~\ref{fig:fig1_1}C).

\subsubsection{Lifetime distribution}
The lifetime distribution of bonds, $P(t)$, is determined from the observed bond lifetimes $h(i,j)$  thatrepresent the continuous duration of a bond between patches $i$ and $j$. The distribution is normalized according to
\begin{equation}
P(t) = \frac{\sum\limits_{ij} \delta(h(ij)-t)}{\sum\limits_{t}\Delta t \sum\limits_{ij} \delta(h(ij)-t)}.
\end{equation}

\section*{Acknowledgments}

This work was supported by the Deutsche Forschungsgemeinschaft (DFG) through SFB 1286 (project ID 317475864), project C05 (to S.K.). This research was conducted within the Max Planck School Matter to Life, supported by the German Federal Ministry of Education and Research (BMBF) in collaboration with the Max Planck Society (A.T., S.K.).

\newpage
\bibliography{simons_temporary_bibliography,bibliography}

\end{document}